# HOW DO THE COVID-19 PREVENTION MEASURES INTERACT WITH SUSTAINABLE DEVELOPMENT GOALS ?

Shima Beigi, Ph.D.[a1*]

## Abstract

Washing hands, social distancing and staying at home are the preventive measures set in place to contain the spread of the COVID-19, a disease caused by SARS-CoV-2. These measures, although straightforward to follow, highlight the tip of an imbalanced socio-economic and socio-technological iceberg. Here, a System Dynamic (SD) model of COVID-19 preventive measures and their correlation with the 17 Sustainable Development Goals (SDGs) is presented. The result demonstrates a better informed view of the COVID-19 vulnerability landscape. This novel qualitative approach refreshes debates on the future of SDGS amid the crisis and provides a powerful mental representation for decision makers to find leverage points that aid in preventing long-term disruptive impacts of this health crisis on people, planet and economy. There is a need for further tailor-made and real-time qualitative and quantitative scientific research to calibrate the criticality of meeting the SDGS targets in different countries according to ongoing lessons learned from this health crisis.

## Introduction

The COVID-19, a disease caused by SARS-CoV-2 started on 12 December 2019 in Wuhan, China. Shortly after, the city of Wuhan became the first epicentre of Covid-19[1]. In an attempt to contain the virus, the Chinese government imposed a 2-month quarantine. However, this initial response did not extinguish the spread of the virus[2]. On 11 March 2020, the World Health Organization (WHO) triggered global health pandemic protocol and Scientific and Technical Advisory Group for Infectious Hazards (STAG-IH) issued a set of public guidelines such as social distancing, washing hands with soap and water and lockdown in the majority of affected areas[3]. As of October, more than 210 countries and territories have reported more than 37 million novel coronavirus cases worldwide (ncov – CSSE). The efforts to contain the virus and slowdown the spread or to flatten the curve continues across the world[4] only to highlight the life-saving importance of the previous global efforts to meet the vision of SDGS[5] and the urgency to create a more resilient world[6].

## Overview

Sustainable development is about bringing balance to the intricate connection between society, economy and environment[7,8]. To achieve such balance, the 17 Sustainable Development Goals (SDGs) are designed[9]. The SDGS envision a unifying roadmap to contribute to the wellbeing of the present population and those to come[10]. SDGS 1-2-3-4-5-6-7-11-16 are social goals, SDGS 13-14-15 are environment goals, and SDG 8-9-10-12-17 are connected to the economy. These goals are intricately interconnected meaning the successful meeting of each depends on the success of the others. The SDGs aim to bring a global vision on human wellbeing beyond linear variables such as gross domestic production (GDP) and average income per household. In this vision, human wellbeing constitutes access to basic materials, fair governments, mental health and freedom of expression. In times of crisis, the value of all these targets become dire and maintaining their continuity becomes even more challenging. Current COVID-19 scientific literature is focused on the unfolding of the crisis and ways to contain it, with few studies on the future of SDGS. The current coronavirus pandemic highlights the urgent need of meeting SDGs 1-2-3-4-6-8-9-10-11-12-13-16-17 **(Figure. 1A).** Pandemics threaten the most vulnerable demographics which lack access to these lifesaving resources. Unpacking the complexity behind these vulnerabilities needs a systemic tool. With the aid of SD[11,12,13], feedback mechanisms and tradeoffs behind the preventive measures and their connection to the SDGS is presented **(Figure. 1B).** The results are valuable for informing debates on the future of SDGs beyond the current crisis.

---

[1a]*Corresponding author. shima.beigi@vub.ac.be, Resilience Scientist, Center for Interdisciplinary Studies, Vrije Universiteit Brussel, Belgium.

**Methodology**

To control the spread of the virus or the so-called flattening of the curve, preventive measures such as washing hands, social distancing and staying at home are put in place. But, what does it take to wash your hands, stay at home and live in lockdown cities? Who are really the most vulnerable groups? A short answer: to wash your hands you need to have access to Safe Water and Sanitation (SDG-6) and to stay at home one needs shelter which implies having a sustaining source of income. A short answer to the second question though, who the most vulnerables are, is not possible as it needs a considerate amount of intellectual, moral and philosophical effort. Here, by using SD, I wish to elaborate on a few of these vulnerabilities and their connections to SDGs. SD concerns how systems change over time and through which pathways[14]. While the approach is heavily based on nonlinear dynamics and system thinking[15], it offers a simple yet extremely powerful method to actively engage with diverse groups of stakeholders. In times of crisis, due to heightened levels of alertness and stress, linear causality is very common. Linear causality focuses on command and control and attempts to solve problems through what is called lock the barn door after the horse is stolen[16]. SD breaks this pattern of thinking through Causal Loop Diagramming (CLD) technique. CLD shows that relationships in complex systems are not linear, may be subject to time lags and feedback loops and therefore are not bounded[17,18]. During a crisis, understanding of feedback loops and time lags is critical. Without having a clear image of reality seeing the forest for the trees is inevitable because decisions tend to be made in isolation and separate from their likely consequences. As a result, recovery from setbacks becomes unnecessarily complex which implies timely allocation of critical resources will suffer. This is particularly evident in the current COVID-19 pandemic where taking short-term measures may be possible[19], but long-term planning is impossible. Both scientific discussions as well as industry discussions on impacts of COVID-19 on people, environment and economy are very fragmented and tend to rely on short-term measures. Scientific publications are overwhelmingly focused on quantitative data, possible early trials of vaccines and other therapies[20,21]. Industry is bustling about COVID-19 impacts on the economy and businesses[22,23]. Besides, a rich picture of the vulnerability landscape of COVID-19, beyond the acute symptoms, which shows existing leverage points is missing. Different stakeholders need to clearly know despite the crisis they can contribute to prevention of this pandemic. Over four months of this crisis is passing, it is time for qualitative approaches, artistic narratives[24] and storytelling methods. These approaches are able to capture the collective imagination over the future of humanity in this generation and those to come. Society, economy and environment are the three main pillars of sustainable development. Yet, these need to be tied together through imagination which is rarely discussed in the sustainability discourses. Without imagination, connecting SDGS and building an inclusive story is impossible. SD brings imagination back into the complex problem space of crises and helps science to be truly in service of all humanity. SD also helps break through public mental barriers in communicating SDGS[25]. Here, a CLD model of COVID-19 preventive measures and their connections with the 17 SDGS is presented (**Figure. 1B).** The model is generated by SD Software VENSIM PLE[26] with a caveat that data on the interconnection of COVID-19 preventive measures with SDGS is very limited with the only available resource provided by the UN[27]. The findings, further elaborated in the discussion, provide a novel perspective on the future of sustainable development research and highlight opportunities for building a global momentum to achieve SDGS targets.

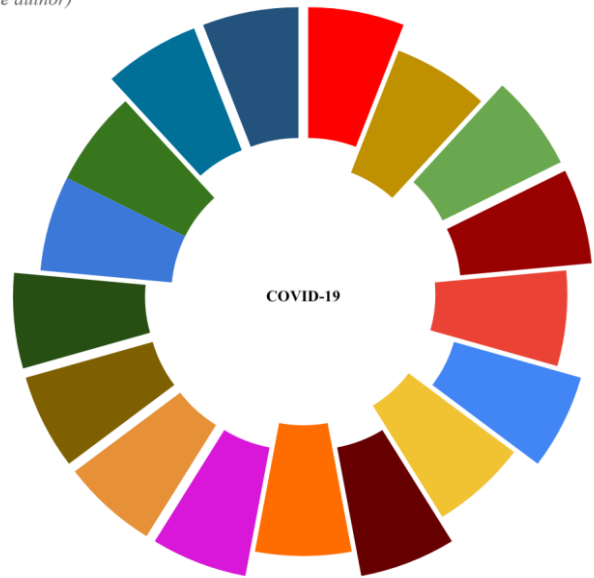

**Figure 1. A. Highlighting the Impact of COVID-19 on the Related Sustainable Development Goals (SDGS).** The current coronavirus pandemic highlights the urgent need to address SDGs 1-3-4-6-8-9-10-11-12-13-16-17 that are shown with distance from the centre.

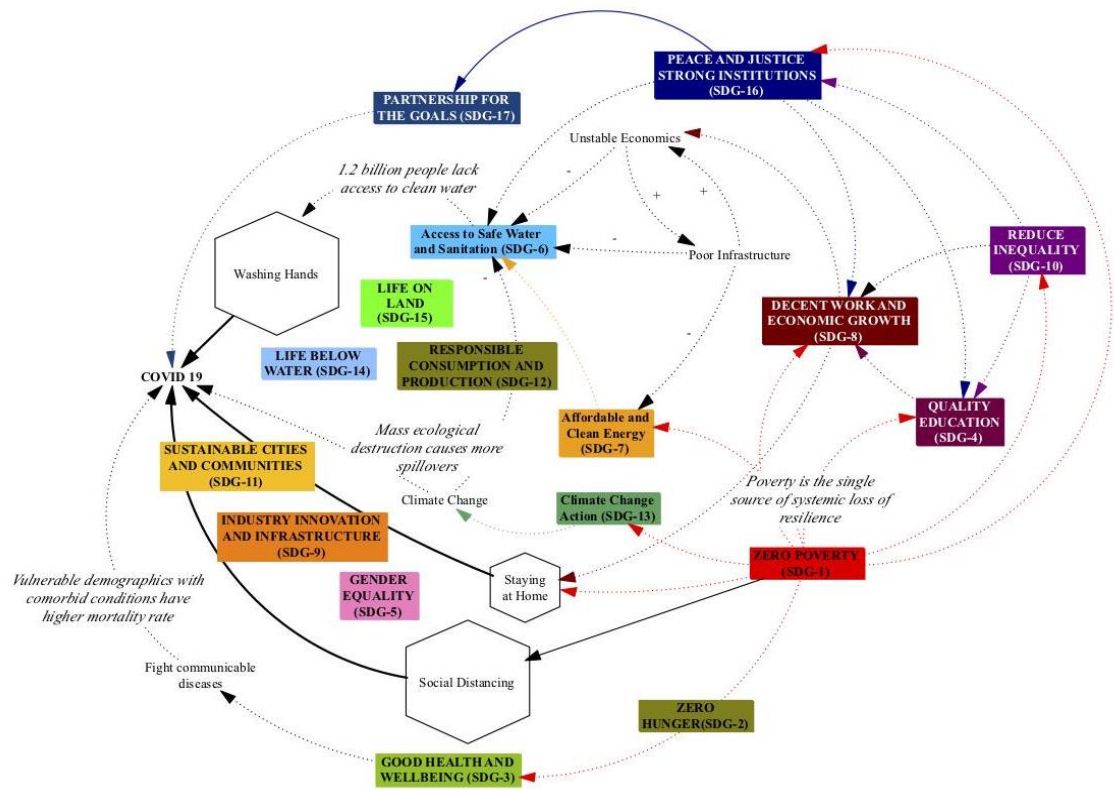

**Figure. 1. B. COVID-19 Preventive Measures (staying home, social distancing, staying at home) and Their Connection to the SDGS.** The Causal Loop Diagram (CLD) shows how health measures set by the WHO through major feedback loops connect to SDGs. Solid arrows show the contribution of WHO preventive measures and dotted arrows track the related SDGS. Poor infrastructure and unstable economies create a vicious cycle that hinder access to safe water and sanitation (SDG-1). Washing hands is also impacted by climate change, one of the main causes of zoonotic spillovers which further highlights the need for urgent efforts to meet the goal of Climate Change Actions (SDG-13). *(Source: Author.)*

**Discussion**

**Access to Clean Water and Sanitation (SDG-6):** To control the spread of the virus or the so-called flattening the curve, preventive measures such as washing hands, social distancing and staying at home are put in place by STAG-IH. Across all major affected regions, COVID-19 immediately threatens the health of vulnerable groups such as the elderly[28] and those with underlying comorbid conditions and with weakened immune systems (i.e., HIV, diabetes, high blood pressure, heart problems)[29]. However, this is only a portion of reality. COVID-19 is epidemiologically different from other SARS-COV[30] in that it is more contagious[31] and therefore adhering to precautionary public health measures is crucial. Yet, these measures should not be taken for granted because they point to the tip of an imbalanced socio-economic and socio-technological iceberg. On the first glance, these are simple guidelines aimed to contain the current health crisis. Nevertheless, by expanding the scope of vulnerability beyond what we learned from medical evidence, a different reality emerges. Take for instance washing hands, according to the United Nations (UN), 3 in 10 people in the world, 1.2 billion people, lack access to safe and readily available water and home. This means 1.2 billion people are not able to follow the basic life saving measures set by STAG-IH to safeguard against COVID-19. These people are mainly located in developing and under-developed countries with impoverished economies and unstable governments as well as countries with extreme weather susceptible to the impacts of climate change. Further, supportive infrastructure systems and technologies are essential operational elements of resilient water infrastructure systems. Due to the long-term impacts of wars, Libya, Lebanon, Syria, Afghanistan, Pakistan, Iraq and neighboring countries[32] are particularly at risk of future pandemics and other crises. In today's language interconnectivity, this means their vulnerability puts others at risk[33]. The combined impacts of these instabilities hinder reaching targets of (SDG-6), a vital prerequisite for washing hands. Frequently washing hands is a short-term preventive measure. But, the social momentum it has created should be harnessed toward a long-term, global effort aimed to reduce global and regional disparities in access to water and sanitation[34].

**Climate Change Actions (SDG-13):** Increased urban interconnection[35], lack of urgency and shared purpose in response to global health crisis and absence of strict codes to address China's wildlife crisis[36] are three main factors that have contributed to the COVID-19 pandemic. Global urban interconnectivity means cities contain so many people, infectious diseases can spread faster and to more people. When 1.2 billion people lack access to clean water (SDG-3), what it really takes to contain the virus is beyond the STAG-IH measures. The real challenge is therefore making access to clean water a priority once for all. This automatically means aiming to meet the Zero Poverty Goal (SDG-1) as soon as possible. Governments should urgently plan to map out creative ways to bridge economic gaps between different social levels[37]. The young generation is eager to rise to these challenges. They only need to be given leadership space. Poverty intervention should aim to identify the population below the poverty line, eradicate undernourishment and provide access to basic health levels, encourage and guarantee maximum average years needed for schooling, based on current world challenges. Global interconnectivity also means aggregated pressures on vital resources which further escalates existing ecological problems of man dominance era. The term used to describe this epoch is called the Anthropocene which refers to a human dominated era marked by fundamental transformation of Earth caused by human affairs[38]. Ecosystems and habitats have been converted to industrial scenes. In addition, urbanisation has become a new human settlement pattern[39] contributing to further pollution, mass consumption and changing Earth's most important evolutionary pressures. To demonstrate safe areas of operation for human development, Earth scientists introduced the model of planetary boundaries[40]. The model shows pressure on natural systems has passed an alarming point[41]. Collapse of ecosystems, urbanisation[42], loss of biodiversity, ocean acidification, and last but not least deforestation contribute to large scale displacement of humans and change in species migration behaviour[43]. As a result of habitat fragmentation, the disconnecting of areas of the landscape from one another, animals are forced to adapt to new environmental conditions[44]. Large scale displacements as a result of interruption of migratory pathways are physiologically taxing to many species, making them vulnerable to different pathogens[45].

This is particularly important to mention in zoonotic circulation such as SARS-related coronaviruses that can be transmitted from animals to humans. While the possible intermediate host(s) of COVID-19 is still unknown[46], the origin of the novel coronavirus is believed to be in bat populations[47]. The family of coronavirus is diverse and can infect animals and humans through an intermediate host such as piglets, birds or camels[48]. Viruses and bacterias have a long history of coevolution and coexistence within the human population. Hence, changes to their adaptation is not alien to humans' immune system. But this co-adaptation dance can be disrupted. With increased levels of ecological degradation and animal displacement, viruses and bacteria are given constant evolutionary advantage to adapt to different hosts, mutate, become more contagious, fatal and jump into the human population[49]. Therefore, the targets of SDG-13 should be recontextualised with the existing scientific evidence[50] on the catastrophic consequences of mass ecological destruction for public health[51]. Maintaining the barrier between the natural reservoir and human society to predict future outbreaks[52] is one of the main management strategies that should be at the forefront of global climate change actions (SDG-13). In addition, the COVID-19 pandemic has to beseech an urgent global climate change action protocol beyond formalities of the Paris Agreement[53,54,55], engage with all industries, businesses, people, and build reliable real-time monitoring models to protect and conserve key ecological systems across the world. We are in this together.

**Zero Poverty (SDG-1), Zero Hunger (SDG-2), Good Health and Wellbeing (SDG-3):** In essence, poverty in all shapes and forms is the mother of most if not all human sufferings. Poverty (absence of SDG-1), lack of access to safe water (absence of SDG-6) and poor health condition (absence of SDG-2 and SDG-3) are only three key vulnerability drivers of pandemics. Meeting Zero Poverty (SDG-1) is a very complex goal. A major lesson of COVID-19 is that eradicating poverty is the most important SDG amongst other SDGS. Poverty directly influences other critical SDGS namely Climate Change Action (SDG-13), Conservation of Life below Water (SDG-14), Life on Land (SDG-15), and Partnership for Goals (SDG-17). Staying at home has become a double-edge sword; saving some while putting a large portion of the population at risk of unemployment (absence of SDG-8) and poverty (absence of SDG-1). In times of crisis managing short term and long-term thinking is crucial. With respect to COVID-19, the short-term focus has been placed on those whose physiological condition put them at risk of developing acute symptoms. Notwithstanding this clinical evidence, in reality, low income workers, young entrepreneurs, women and young children in poor areas are at greater long-term risk. Chronic exposure to stressful situations overloads the body and leads to loss of neurological, physiological and psychological resilience (i.e., Loss of SDG-3). Had the poverty gaps in major COVID-19 epicentre been decreased prior to the outbreak, the recovery phase would have been smoother and less financially burdensome. Another short-term focus is placed on the COVID-19 vaccine[27]. While vaccines are key to the future efforts to contain the virus, they are unlikely to cure the root cause of crises. Therefore, creating a positive synergy between nations to urgently address these vulnerabilities and prevent future crises should be given equal resources and global attention.

**Partnership for the Goals (SDG-17):** The COVID-19 pandemic shows mobilisation of resources towards a sustainable world vision has to be prioritised on poverty, health, education, climate change and better and transparent governance systems. Nonetheless, one of the main obstacles to make this shift is the silo-like operation of policy makers and governance systems around the world and lack of stakeholders' awareness of SDGS operationalisation. Mutually reinforcing efforts across SDGs require a shared understanding that SDGs intricately depend on each other. To meet the requirements of SDG-17, clustering SDGs according to priorities and needs of local and global communities is one possibility. Governance as a critical pillar of sustainable development is more potent when it is polycentric and engages with a diverse group of stakeholders[56,57,58,59]. With the aid of sustainable friendly artificial intelligence (AI)[60], clustering of the SDGS can trigger emergence of vectors of international polycentric engagements to meet the vision of sustainable development. Attempts to address the urgent need for Partnerships for the Goals (SDG-17) is further raised by the United Nations Secretary Antonio Guterres calling for an immediate ceasefire in all corners of the world to focus on the global fight against COVID-19. COVID-19 can become a tale of SDG tragedy. However, it could also be turned into a tale of triumph. To do that, partnership should focus on building democtaric platforms that allow sharing of critical information regarding the progress of SDGs across the world with or without crises.

**Gender Equality (SDG-5), Industry Innovation and Infrastructure (SDG-9), Reducing Inequality (SDG-10):** Reducing inequality is often seen in terms of income and GDP. However, inequality akin to poverty is caused by a combination of mental and physical factors. Lack of digital education and access to valuable sources of

Preprints (www.preprints.org)

information are amongst recent challenges humanity is facing. False information and the rise of fake news pray on these underlying vulnerabilities particularly in crises. Social-media companies must become more transparent, accountable and socially beneficial[61]. With the lockdown being imposed on almost half of the world population, Industry, Innovation and Infrastructure (SDG-9) is directly impacted. The demand on digital platforms and information communication technologies (ICT) is increasing. Many schools, universities and educational institutions have responded to this challenge and became digital, reaching students and learners remotely. This transition occurs whilst many families do not have access to the necessary technological skills and knowledge to rise to the challenge. Put it simply, access to the internet does not mean internet skills. The impacts of social distancing and lockdown around the world should be put into the bigger context and connect with digital inequality and its impacts on socio-economically disadvantaged groups[62]. The staggering demand on ICT platforms should make it crucial to establish a universal digital education plan as a part of reducing inequality (SDG-10). The digital gap is particularly noticeable in girls and women whose role as unpaid caregivers is often taken for granted[63]. With the majority of world ICT platforms being male dominated, Gender Equality (SDG-5) should be seen beyond sex differences and updated with regard to technological and digital challenges women face around the world. Governments should dedicate necessary social, educational and cultural resources to women entrepreneurs. Giant technology developers must rise to the challenge of demotrasticing (AI)[64] and connection to sustainable development and actively play a role in encouraging a more gender balanced, mindful and responsible digital ecosystem[65].

**Sustainable Cities and Communities (SDG-11):** SDG-11 aim is to "*... substantially increase the number of cities and human settlements adopting and implementing integrated policies and plans towards inclusion, resource efficiency, mitigation and adaptation to climate change, resilience to disasters, and develop and implement, in line with the Sendai Framework for Disaster Risk Reduction 2015-2030, holistic disaster risk management at all levels by 2020."* Today, we are in 2020 facing a health crisis that is devastating cities. As the crisis unfolds, cities will be overwhelmed by COVID-19 which means efforts to meet SDG-11 must be redirected to address these concerns. This is crucial because urbanisation is rapidly growing. By 2050, more than 65% of the world population, close to 7 billion people, will live in cities[66] **(Figure. 2).** The economic role of cities is significant as they generate about 80 percent of the global GDP. The coronavirus pandemic has taken a halt on the activities of most dense urban areas across the world.

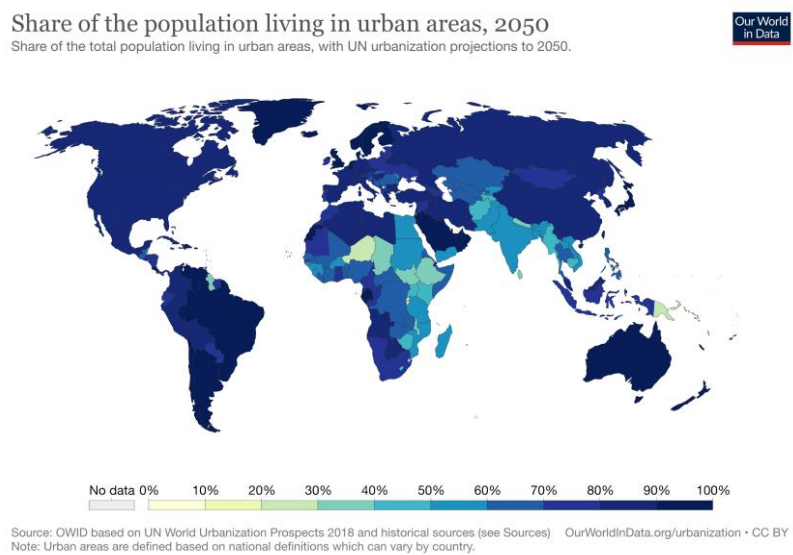

**Figure 2. Share of the Population Living in Urban Areas by 2050.** *(Source : UN HABITAT)*

During the lockdown, many cities are forced to mobilise their economic resources to protect citizens from the virus. Social distancing and staying at home means further demand for vital infrastructure services. Add to this challenge, the impact of climate change on vulnerable urban systems. 1-in-3 urban dwellers live in slum households with aging or deteriorated infrastructure systems **(Figure. 3).**

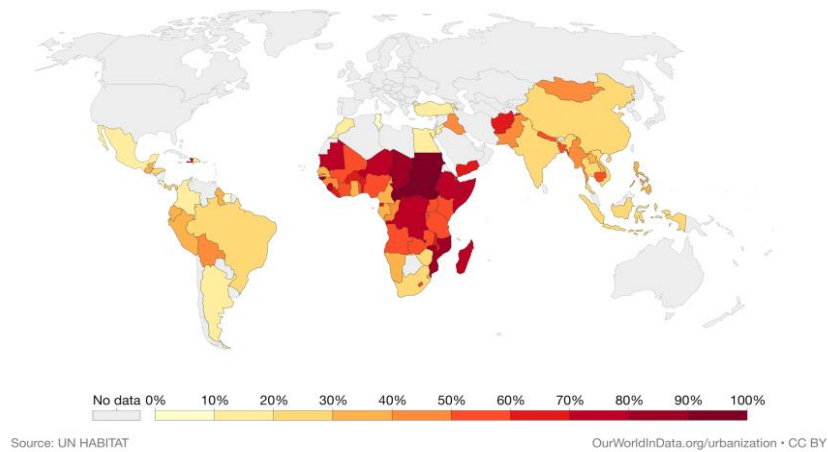

**Figure 3. Share of Urban Population Living in Slums.** *(Source: UN HABITAT)*

If the urban unsustainable development trajectory is not publicly and scientifically debated[67, 68], expecting more dangerous and frequent epidemics is inevitable. Crisis further lay bare existing vulnerabilities. As if the US sanctions were not sufficient to bring the Iranian economy to total economic lockdown[69], amid the COVID-19 crisis, the south and central provinces of Iran were hit by spring floods causing damage to 9,934 kilometres of roads and 886 bridges[70]. These provinces are all located in highly impoverished areas of the country with fragile infrastructures and poor access and contribution to the national economy **(Figure. 4).**

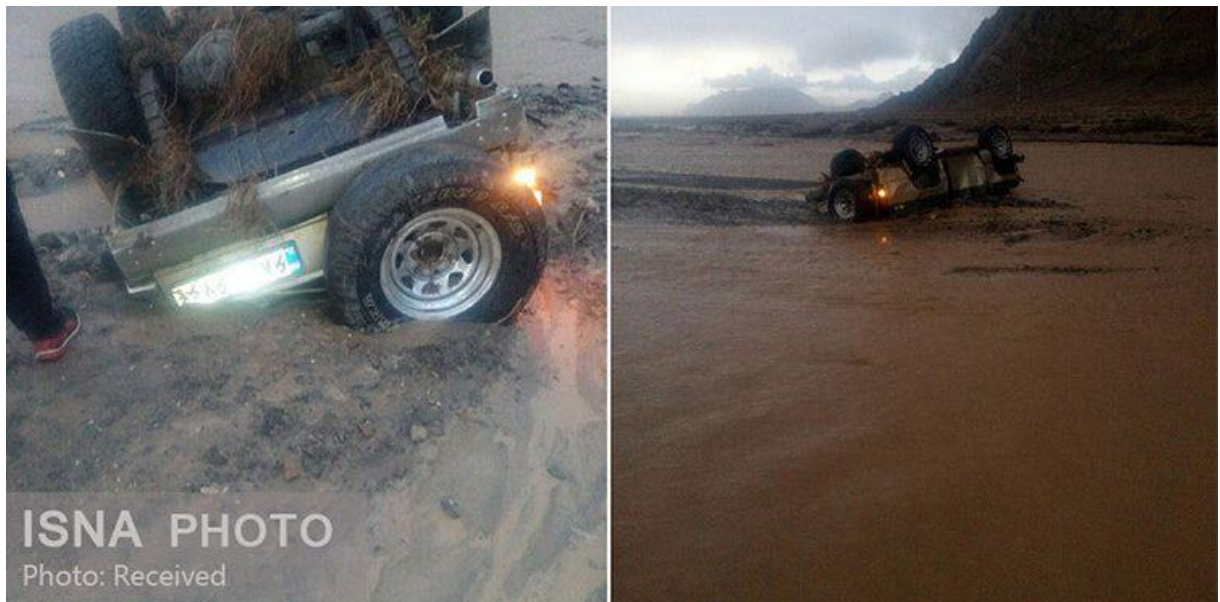

**Figure 4. Seasonal rains in 16 Iran provinces during COVID-19.** *(Source: Iran News Wire)*

Iran as one of the hardest hit countries by COVID-19 is only one example of countries that are in danger of catastrophic impacts of climate change[71]. Highly important socio-cultural cities in Italy, several metropolitan states in the USA, highly touristic spots in France and Spain and many other large socio-economic hubs exhibit various catastrophic impacts of the coronavirus health crisis on the fabric of urban systems. The data from the european aviation control centre (EUROCONTROL) shows a staggering 89% fall in travel across european cities compared to April 2019 ([Daily Traffic Variation - States](#)). Many of the affected european cities draw income from tourism, loss of which leads to alarming socio-economic impacts. Furthermore, European cities were already struggling to meet the demand of immigration crisis[72], governance and economic crises[73]. Addressing these challenges is urgent. Before the coronavirus pandemic, cities were considered as melting pot[74] and scored according to their economic performance[75] and capacities to attract the most talented people. The reports of the World Urbanisation Prospects Report (United Nations) usually projected the future of urbanisation without taking into account vulnerability indices of cities that are in danger of future social, technological, health or political crises. Further, an overwhelming amount of urban scientific effort is focused on the economic power and growth of cities. However, to survive cities need to constantly reinvent themselves[76]. Crisis not only demands cities to be prepared for disaster, but also calls for the triggering of new innovation cycles. Nonetheless, COVID-19 shows even the most economically booming cities are extremely fragile in setting off these cycles. A solution is an urgent urban sustainability investment vision[77] to empower citizens as the main driver of creativity, protect ecosystem services and cultivate a culture of peace[78]. Equally important, it is the role of green and open spaces to citizens' wellbeing[79]. Previous efforts to link the risk of urban public health to contagious diseases[80,81] should be pushed towards a systemic placed-based mechanism of urban welfare assessment based on the requirements of Good Health and Wellbeing (SDG-3) that ensures public health and psychological wellbeing is included in construction and engineering, urban planning, smart cities, urban policy and governance.

**Conclusion**

Under the pressing challenge of COVID-19 the future of Sustainable Development is debatable. What constitutes a sustainable development vision varies from one place to another. This is particularly noticeable when the preventive measures set by the WHO (i.e., washing hands, social distancing and lockdown ) are linked to the 17 SDGS. Washing hands should not be taken for granted as a basic strategy for it is implicitly connected with access to safe water and sanitation or SDG-3. Social distancing and staying at home are most effective when the supportive resources are or made available. Social distancing directly calls on SDGs 5, 9, and 11. With Zero Poverty (SDG-1) being the most important common denominator in flattening the curve. Social distancing aims to prevent increase of contagion among the population and regulate demands on health infrastructure systems. It is crucial to understand that these preventive measures are short-termed and will go only so far. Meaning they cannot provide long-term sustainable resolutions. What needs to be done is connecting these efforts to the existing agenda of SDGS and related goals such as Zero Hunger (SDG-2) and Good Health and Wellbeing (SDG-3). The role of AI, technology, innovation and infrastructure systems in supporting global efforts should not be taken for granted. Further funding needs to be allocated to the role of Industry, Innovation and Infrastructure (SDG-9) in shaping the future of crisis response. During most crises, creative use of technology specially AI has shown promise. Technology companies must therefore rise to these challenges and foster a more democtaric and gender diverse ecosystem. Overall, COVID-19 shows that there is a need for a tailor-made and real-time qualitative and quantitative scoring mechanism to prioritise SDGS in different countries according to lessons learned from this health crisis.